# Cometary airbursts and atmospheric chemistry: Tunguska and a candidate Younger Dryas event.

Adrian L. Melott[1], Brian C. Thomas[2], Gisela Dreschhoff[1], and Carey K. Johnson[3]

[1]Department of Physics and Astronomy, University of Kansas, Lawrence, Kansas 66045

[2]Department of Physics and Astronomy, Washburn University, Topeka, Kansas 66621

[3]Department of Chemistry, University of Kansas, Lawrence, Kansas 66045

**ABSTRACT**

We find agreement between models of atmospheric chemistry changes from ionization for the 1908 Tunguska airburst event and nitrate enhancement in GISP2H and GISP2 ice cores, plus an unexplained ammonium spike. We then consider a candidate cometary impact at the Younger Dryas onset (YD). The large estimated $NO_x$ production and $O_3$ depletion are beyond accurate extrapolation, but the ice core peak is much lower, possibly because of insufficient sampling resolution.

Ammonium and nitrate spikes have been attributed to biomass burning at YD onset in both GRIP and GISP2 ice cores. A similar result is well-resolved in Tunguska ice core data, but that forest fire was far too small to account for this. Direct input of ammonia from a comet into the atmosphere is adequate for YD ice core data, but not Tunguska data.

An analog of the Haber process with hydrogen contributed by cometary or surface water, atmospheric nitrogen, high pressures, and possibly catalytic iron from a comet could in principle produce ammonia, accounting for the peaks in both data sets.

**INTRODUCTION**

Firestone et al. (2007) suggested a North American comet airburst as trigger for climate change/mass extinction at the Younger Dryas (YD) onset 12.9 kya. A bolide ionizes the atmosphere (Reid and McAfee 1978), breaking the $N_2$ triple bond to enable synthesis of (normally low abundance) oxides of nitrogen. We previously computed estimates which agree reasonably with ice core data on nitrate deposition at the largest recorded solar flare (Thomas et al., 2007). We now test this method on the Tunguska bolide event of 1908, then extend it to the candidate YD impactor.

We estimate the masses of the objects as $5 \times 10^7$ kg for Tunguska (Wasson, 2003), and $5 \times 10^{13}$ kg for YD (Firestone et al., 2007). Recent work suggests that the Tunguska object was likely a comet or fragment (Kelley et al., 2009). We assume half cometary ice and half rock for both events, but this matters little for atmospheric ionization. We assume a speed of ~30 km s$^{-1}$ (Steel, 1998), the typical orbital velocity of objects bound to the Sun. This corresponds to ~10 eV of kinetic energy per proton, easily producing temperatures ~$10^5$ °K in the vicinity of the impact. The $N_2$ dissociation energy is ~10 eV (Frost and McDowell, 1956), producing monatomic nitrogen which will then react to produce $NO_x$ (NO, $NO_2$) and $O_3$ depletion. Given the aerial burst assumption for both events, we have $4 \times 10^{16}$ J and $4 \times 10^{22}$ J, respectively for these events.

**$NO_x$ SYNTHESIS, $O_3$ DEPLETION, AND NITRATE DEPOSITION IN ICE CORES**
**Computational Methods**

We have used the Goddard Space Flight Center (GSFC) 2D atmospheric model to estimate nitrate ($NO_3^-$) deposition following the 1908 Tunguska event. The model has been used extensively in the past to model various effects of ionizing photon and charged particle events, from such events as nearby supernovae (Gehrels et al. 2003), gamma-ray bursts (Thomas et al., 2005; Melott and Thomas 2009) and major solar flares (Thomas et al., 2007). The model computes 65 constituents, with photochemical reactions, solar radiation variations, and transport, with a one-day time step. More details may be found in Thomas et al. (2005; 2007), and references therein.

While the impactor only initially affected a small area, the atmosphere is well mixed longitudinally in 10 days or so. We divide the total energy by the area of the 65° North latitude band for the Tunguska event (~2 × 10$^{13}$ m$^2$) and by 10 days. We assume that there is one ion pair created for every 35 eV of input energy (Porter et al., 1976), giving 4 × 10$^{14}$ ions m$^{-2}$ s$^{-1}$. This is a source of NO$_x$, assuming 1.25 NO$_x$ molecules are produced per ion pair (Porter et al., 1976). We distribute the ionization in altitude between 10 and 50 km (Park, 1978), keeping the ratio of ionization/N$_2$ constant with altitude. The input is a 10 day step function starting on 30 June.

One can use scaling relations to make estimates outside the range of validity of our numerical methods. In past work, NO$_x$ and nitrate production scale as input energy E (Thomas et al., 2005). Percentage O$_3$ depletion scales ~$E^{1/3}$. Both are independent of the input timescale, for times ranging from 100 ms to 3 yr (Ejzak et al., 2007).

**The Tunguska Event**

The GSFC model computes surface nitrate deposition density at each time step. We compare modeled nitrate deposition to ice core data. Using high-resolution GISP2H ice core data (Summit, at the peak of the Greenland ice cap at 72°34'44.10"N 38°27'34.56"W) from the CODIAC archive at NCAR/EOL (Dreschhoff and Zeller, 1994; Dreschhoff, 2002) we find the enhancement over background for Tunguska. We first sum nitrate measurements at samples 2225–2227 (the associated peak) and then subtract background, which we assume is given by an average of 5 measurements preceding and following the three peak values. We find a net peak enhancement of 165 ppb. This is converted to a surface density by multiplying by the density and thickness of the ice/firn (600 kg m$^{-3}$) giving 1.49 × 10$^6$ kg m$^{-2}$.

To compute an equivalent value from the simulation results, we sum the rate at the 75° North band over 90 days after the event, which gives a net enhancement of ~1.50 × 10$^6$ kg m$^{-2}$, in good agreement with the ice core result.

Similar values were computed for the 1859 Carrington "white light" Solar flare, which appears in the GISP2H ice core data (Thomas et al., 2007). Following the procedure above, we sum the nitrate measurements at sample numbers 3143–3145, yielding an enhancement of 143 ppb and a resulting surface density 1.50 × 10$^6$ kg m$^{-2}$. From our 2007 results modeling the Carrington event and the same procedure described above, we compute an enhancement of 1.52 × $10^6$ kg $m^{-2}$, in agreement with the ice core result.

$NO_x$ destroys ozone in a well-known catalytic cycle. For Tunguska, our modeling indicates a maximum $O_3$ column density loss of ~3% at the 65° North latitude band, with a global average depletion of 0.25% for about a month. Our modeling would suggest no measurable UVB effect on biota as a consequence of Tunguska.

**The Younger Dryas Event**

Our computational model is not capable of handling the much larger energy input value associated with the candidate YD event. However, it is possible to use scaling relations described earlier to make a rough estimate. This suggests the presence of an ice core nitrate signal for the YD event that is $10^6$ times larger than that of the Tunguska event.

The nitrate enhancement that we found in our modeling of the Tunguska event, and also in the GISP2H ice core data, is of order $10^6$ kg $m^{-2}$, which suggests an expected YD ice core nitrate of order 1 kg $m^{-2}$. These scaling relations ignore limiting processes that will come into play at large inputs. There should nevertheless be a significant nitrate signal for this event, well above previously noted spikes. Since transport across the equator is inefficient, the signal for a North American impact should be much greater in Greenland cores than Antarctic cores. We also mention that changes in atmospheric circulations may compromise these results (Fuhrer et al. 1996).

Figures 1a-1c show ice core data for time periods of interest. The nitrate spikes are clear (especially in GISP2) but not large. Existing insufficient stratigraphic (age) resolution is probably an important constraint and a larger spike may have been missed.

It is also possible use a scaling relation to estimate ozone depletion from the YD event. The scaling was determined in the energy range up to ~$10^{20}$ J and does not directly apply at much higher energies. It yields greater than 100% depletion in polar regions, and a global mean of order 30%. We can at the very least predict severe ozone depletion associated with the candidate YD comet impact over the northern hemisphere. Some likely consequences of the UVB enhancement on biota resulting from such ozone depletion are described in Melott and Thomas (2009).

**AMMONIUM ENHANCEMENTS ASSOCIATED WITH POSSIBLE COMET STRIKES**
**An Ammonium Signal Coincident with the Nitrate Signal**

Mayewski et al. (1993) noted an ammonium increase in Greenland ice cores at YD onset corresponding to an average deposition rate of $\sim 10^6$ kg m$^{-2}$ yr$^{-1}$. The sampling resolution through this period is 3.48 yr. Given the ice core resolution and the possibility of diffusion and/or mixing through ice, the integrated values of these quantities are probably more reliable, suggesting that $10^5$ to $10^4$ kg m$^{-2}$ total deposition is required to account for the data. Similar increases also occur for nitrate, though they represent much less of a fractional enhancement.

In Figures 1a-1c we show the time development of the concentrations of $NH_4^+$ and $NO_3^-$ in GRIP (DeAngelis et al. 1997), GISP2 (Mayewski et al. 1993), and an unpublished high-resolution version of GISP2 for recent times (P. Mayewski, personal commun.) for the YD (Figs. 1a-1b) and the Tunguska (Fig. 1c) events. We show these records in order of increasing time resolution. The difference in onset time between Figures 1a and 1b corresponding to the highest peak reflects the well-known dating offset of up to several hundred years for YD onset between the two data sets (Southon 2002; Rasmussen et al. 2006; Vinther et al. 2006). The main item of interest here is the time development of the simultaneous maxima in the two ions, suggesting a common cause. In all three cases the absolute enhancement of $NH_4^+$  $NO_3^-$. We now consider possible production mechanisms. Atmospheric ionization is not known to produce ammonium (Thomas et al. 2005).

The removal of nitrate from the atmosphere has a characteristic timescale of a few years, comparable to the sampling interval for Figure 1b, more rapid for polar regions (Thomas et al. 2005, 2007). If YD climate change produced a temporary pulse of increased rainfall, the removal could have been even more rapid (e.g., Haynes 2008) and in addition also leading to a diluted signal, complicating the picture even more. The atmospheric residence time of ammonia is not well constrained, but almost certainly shorter than nitrate (Tsunogai and Ikeuchi 1968).

**Production by Biomass Burning**

The YD spike of ammonium and nitrate was interpreted by Firestone et al. (2007) as an effect of biomass burning (e.g., Schlesinger 1997). The $NH_4^+$ is approximately tripled, an increase of ~49 ppb in the ice at 12,812 BP, noted by Mayewski et al. as the largest such event in the last 110,000 yr. Allen West (personal commun.) notes the sum of many fire indicators reaching a peak greater than any in the preceding 370,000 yr in GRIP, and 100,000 yr in GISP2. Typical biomass burning produces conversion of ~0.1% of the biomass to ammonium and ~0.4% to nitrate (Hegg et al. 1988, 1990; Schlesinger 1997). Boreal forest burning, however, is different, and typically produces more ammonium than nitrate (Lebel et al., 1991). Boreal forests of the type that cover Siberia in the Tunguska region were widespread in North America as well, and typically contain of order 80 Mg/ha of biomass (Houghton et al. 2007). Burning a forest area below the North American ice line, crudely approximated as $8 \times 10^8$ ha, would contribute as much as $10^{-3}$ kg m$^{-2}$ of ammonium distributed over the Northern Hemisphere, and a comparable amount of nitrate, which is comfortably much larger than the ice core signal. Assumption of complete burning is clearly an overestimate, because wildfires typically do not lead to complete burning even locally. Figures 1a-1b show that the increment of ammonium is comparable to that of nitrate. As the Siberian Taiga and much of North America at the time of the YD were covered by boreal forest, the observed mix should not present a problem.

A significant problem exists in using biomass burning to explain the Tunguska event. The synchronous increase in both ions in the ice core data for the winter of 1908–09 in the GISP2 signal is clear and reliably dated with high time resolution. As we shall see, biomass burning can only be a minor contributor to the signal for this event. The area of forest fire burning was only 10–20 km in diameter (Wasson 2003). If we generously assume 100,000 ha of forest burning, the surface density deposition of ammonium over the Northern Hemisphere is only $10^{-7}$ kg m$^{-2}$, only somewhat larger for nitrate. Summing up the nitrate from either GISP2H or GISP2 produces ~$5 \times 10^{-6}$ kg m$^{-2}$ for the nitrate or ammonium deposition in the wake of Tunguska, so clearly the biomass burning is insufficient. The strong signal in the wake of Tunguska is one of the highest peaks over a recent century (Dreschhoff 2002; Olivier et al. 2006).

**Production by Direct Deposition from a Comet**

Oblique impacts can allow considerable survival of even complex molecules (Schultz and Gault 1991; Edwards et al. 2009). We assume comet ice composed of 1% ammonia (Bird et al., 1997; Kawakita and Watanabe, 2002). For YD, this gives ~2 X$10^4$ kg m$^{-2}$ if deposited uniformly across the Northern Hemisphere of the Earth, in agreement with ice core data. However, for Tunguska, this mechanism fails by 5 orders of magnitude to provide enough to account for the ice core data (Fig 1c, based on P. Mayewski, personal communication).

Thus, all three candidate processes fail to explain the atmospheric chemistry record for the recent, and best understood, Tunguska event. Occam's Razor directs us toward a consistent explanation, if the events have similar causation.

**An Alternative: The Haber Process?**

The Haber process for ammonia synthesis was developed for fertilizer and munitions in 1909. Under conditions of high pressure, nitrogen and hydrogen react to form ammonia. Formation of ammonia is increasingly disfavored thermodynamically at higher temperatures with respect to molecular nitrogen and hydrogen because of the unfavorable entropy of reaction. However, the unfavorable free energy of reaction can be overcome by the high pressure present in the shock front of a comet entering the atmosphere. As the nitrates estimated for both the Tunguska and YD from conventional atmospheric process are adequate to explain the data for both events, it is reasonable that the comparable amount of ammonium found in the cores could also be synthesized this way, using cometary ice.

The assumed ice mass in the Tunguska comet is insufficient to synthesize sufficient ammonia to account for Greenland ice cores. However, it was proposed that at least one fragment may have impacted a swampy, partially melted permafrost area, creating Lake Cheko (Gasperini et al., 2008). Based on the size of the probable crater lake, sufficient water would have been present as a reactant to synthesize the ammonia. The water budget is not an issue for the proposed YD event, as either cometary ice mass or interaction with the Laurentide ice sheet would have been sufficient.

**DISCUSSION**

If a major bolide impact is involved, existing ice cores should show a large nitrate signal at the onset of the YD. Such an impact could not take place without the production of large amounts of nitrate. The GRIP and GISP2 data show only a modest enhancement. The expected value is far above background and should stand out. This is a clear prediction of such a large atmospheric event. However, as noted by Mayewski et al. (1993), distribution can be uneven.

The characteristic global timescale for complete atmospheric removal of $NO_x$ under present conditions is of order a few years, (Thomas et al., 2005; 2007), and even shorter for ammonium. The GISP2 ice core had a mean time interval of 3.5 yr between samples (Mayewski et al., 1993). The interval for GRIP is much larger. It is possible, especially since deposition rates are likely to change at the onset of major climate changes, that a nitrate spike may not have been sampled.

Both ammonium and $NO_x$ are produced in biomass burning, and they share a short residence time in the atmosphere (Schlesinger, 1997). Biomass burning widespread across the ice-free portions of North America appears to be consistent with the data for YD.. However, they are also both observed with coincident peaks for the Tunguska event, and the area burned then is far too small to account for the ice core data.

The large ozone depletion expected from such an event as the hypothesized YD comet would definitely cause a UVB crisis in the northern hemisphere, which could contribute to extinction in a variety of ways. Some might be detectable, but discussion is beyond the scope of this study; see Melott and Thomas 2009. Higher-resolution examination of Greenland ice cores at the YD onset is an appropriate and strong test of the cometary impact hypothesis. Examination at higher temporal resolution for higher peaks in nitrate, ammonium, and possible pollen morphology changes (Yeloff et al. 2008, and references therein) should be undertaken.

Given observed similarities in the chemical signature in ice cores, we seek a viable common cause for the two events. We summarize in Table 1 the status of various possible mechanisms for the production of the ions observed in the (probable comet) Tunguska event, and the YD event hypothesized as a comet.

|  | BIOMASS BURNING | | DIRECT DEPOSIT | | ATMOSPHERIC IONIZATION | | ICE + ATMOSPHERIC IONIZATION + HABER | |
|---|---|---|---|---|---|---|---|---|
|  | YD | Tunguska | YD | Tunguska | YD | Tunguska | YD | Tunguska |
| $NO_3^-$ | Adequate | No | Unknown | No | Adequate | Adequate | Adequate | Adequate |
| $NH_4^+$ | Adequate | No | Adequate | No | No | No | Adequate? | Adequate? |

As can be seen, if we demand consistency, the Tunguska event rules out three of the four candidate mechanisms, as the bolide and its ensuing fire were both too small to produce adequate amounts except for nitrate by atmospheric ionization, but this does not normally produce ammonium. Sufficient ionization energy should exist in both events to produce adequate amounts of both species, provided that a Haber-like process including cometary or surface water diverts a substantial fraction of the reaction products into ammonia. Our estimates suggest that there should be greater deposition of nitrate than so far observed from an atmospheric ionization process if the YD event were a cometary airburst of the requisite size. We recommend high-resolution examination of ice core data from the time.

**ACKNOWLEDGMENTS**

GISP2H ice core data was retrieved from the CODIAC archive at NCAR/EOL. The authors thank P. Mayewski for providing additional ice core data, and two anonymous referees for helpful comments on presentation. We gratefully acknowledge research support from NASA grant NNX09AM85G.

Mayewski, P.A., Meeker, L.D., Whitlow, S., Twickler, M.S., Morrison, M.C., Alley, R.B., Bloomfield, P., and Taylor, K., 1993, The Atmosphere During the Younger Dryas: Science, v. 261, p. 195–197, doi: 10.1126/science.261.5118.195.

Melott, A.L., and Thomas, B.C., 2009, Late Ordovician geographic patterns of extinction compared with simulations of astrophysical ionizing radiation damage: Paleobiology, v. 35, p. 311–320, doi: 10.1666/0094-8373-35.3.311.

Olivier, S., Blaser, C., Brütsch, S., Frolova, N., Gäggeler, H.W., Henderson, K.A., Palmer, A.S., Papina, T., and Schwikowski, M., 2006, Temporal variations of mineral dust, biogenic tracers, and anthropogenic species during the past two centuries from Belukha ice core, Siberian Altai: Journal of Geophysical Research, v. 111, p. D05309, doi: 10.1029/2005JD005830.

Park, C., 1978, Nitric oxide production by Tunguska meteor: Acta Astronomica, v. 5, p. 523–542, doi: 10.1016/0094-5765(78)90082-6.

Porter, H.S., Jackman, C.H., and Green, A.E.S., 1976, Efficiencies for production of atomic nitrogen and oxygen by relativistic proton impact in air: The Journal of Chemical Physics, v. 65, p. 154–167, doi: 10.1063/1.432812.

Rasmussen, S.O., and 15 others, 2006, A new Greenland ice core chronology for the last glacial termination: Journal of Geophysical Research, v. 111, p. D06102, doi: 10.1029/2005JD006079.

Reid, G.C., and McAfee, J.R., Jr., 1978, Effects of intense stratospheric ionisation events: Nature, v. 275, p. 489–492, doi: 10.1038/275489a0.

Schlesinger, W.H., 1997, Biogeochemistry: An Analysis of Global Change: Elsevier, San Diego. P. 383–396.

Schultz, P.H., and Gault, D.E., 1991, Impact Decapitation from Laboratory to Basin Scales: Abstracts of the Lunar and Planetary Science Conference, v. 22, p. 1195–1196.

Southon, J., 2002, A First Step to Reconciling the GRIP and GISP2 Ice-Core Chronologies, 0–14,500 yr B.P: Quaternary Research, v. 52, p. 32–37.

Steel, D., 1998, Distributions and moments of asteroid and comet impact speeds upon the Earth and Mars: Planetary and Space Science, v. 46, p. 473–478, doi: 10.1016/S0032-0633(97)00232-8.
Page 12 of 14

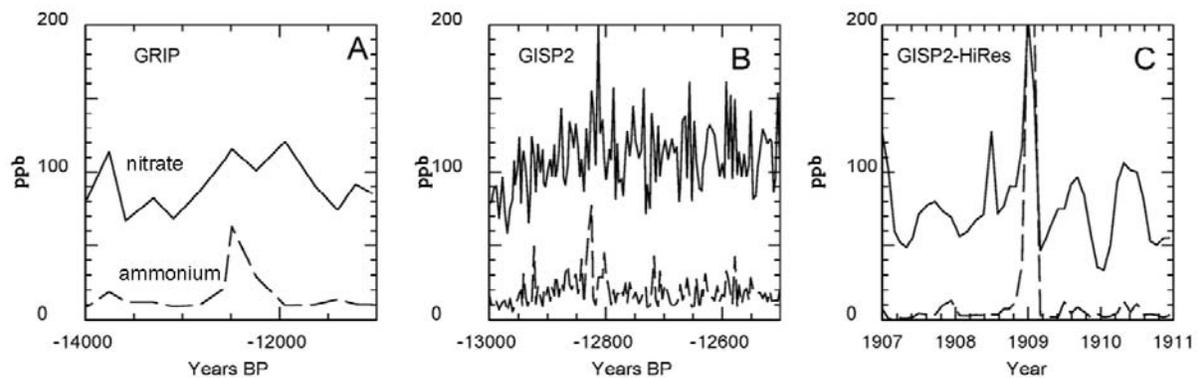

**FIGURE CAPTIONS**

Figure 1. Time development of ammonium and nitrate ions in two samples taken at YD onset and at the time of the Tunguska event. Note that there is known age offset between GRIP and GISP2, described in the text. A: Concentration of nitrate (solid line) and ammonium (dashed line) ions in the GRIP core. YD onset event is in the ice core dated 12491 BP. Time resolution of this sampling is low, compared to the expected brief atmospheric residence time of these species. Nevertheless, the values are remarkable, and larger than prior peaks for 386,000 yr. 1B: Concentrations in the GISP2 core. YD onset event is in the ice core dated 12812 BP. Note the major peak at the YD onset. Time resolution of this core is comparable to the probable global average residence time of nitrate in the atmosphere under present conditions; ammonia residence time is shorter. 1C: Concentrations in the GISP2 high-resolution core. The Tunguska event, thought to be a comet airburst, took place in 1908. Time resolution of this core is much better than the residence time of nitrate in the atmosphere.